\documentclass[]{revtex4}
\usepackage[dvips]{graphicx}
\def\gsim{\buildrel {\textstyle >}\over {_\sim}}
\def\lsim{\buildrel {\textstyle <}\over {_\sim}}
\begin{document}

\title{
Spin and chirality orderings of the one-dimensional Heisenberg spin glass with the long-range power-law interaction
}

\author{Akihiro Matsuda, Mitsuru Nakamura and Hikaru Kawamura}

\address{Faculty of Science, Osaka University, Toyonaka 560-0043, Japan}
\begin{abstract}
The ordering of the one-dimensional Heisenberg spin glass interacting via the long-range power-law interaction is studied by Monte Carlo simulations.  Particular attention is paid to the possible occurrence of the ``spin-chirality decoupling'' for appropriate values of the power-law exponent $\sigma$. Our result suggests that, for intermediate values of $\sigma$, the chiral-glass order occurs at finite temperatures while the standard spin-glass order occurs only at zero temperature.

\end{abstract}

\maketitle

\section{Introduction}

In order to understand the true nature of the experimental spin glass (SG) 
transition, particularly of canonical SGs which possess nearly isotopic 
interaction, it is crucially important to elucidate the ordering properties of
the three-dimensional (3D) isotropic Heisenberg SG. Some time ago, one of the
present authors (H.K.) proposed that the 3D isotropic Heisenberg SG might
exhibit an intriguing ``spin-chirality decoupling'' phenomenon
at long length and time scales,
{\it i.e.\/}, the ordering of the chirality occurred at a temperature higher
than the standard SG transition temperature. Chirality is a multispin variable representing the sense or the handedness of the noncoplanar spin structure induced by spin frustration. 
It was suggested that such a spin-chirality decoupling might
play a crucial role in the experimental SG ordering \cite{Kawamura92,Kawamura96}. 

Concerning the possible occurrence of such spin-chirality decoupling in the 3D Heisenberg SG, however, controversy has continued some time now. Different numerical simulations
by different authors reported apparently opposite conclusions \cite{HukuKawa,HukuKawa2,Matsubara,Nakamura,LY,Matsubara3,BY,Picco,Campos}.
Difficulty in finite-size numerical simulations might lie in the fact
that, as emphasized in Ref.\cite{HukuKawa2}, the spin chirality-decoupling,
if any, is realized only at longer length scale beyond a
crossover length $L^*$, so that one needs to go beyond this crossover
length in order to really see  whether the spin-chirality decoupling occurs or not. 

To understand the issue in wider perspective, it might be useful to study the phenomena by generalizing the dimensionality $d$ from original $d=3$ to both lower and higher dimensions. In the limit of high dimension $d\rightarrow \infty$, the model reduces to the mean-field (MF) Heisenberg SG model. In the MF limit, it has been known that there is a single SG transition at a finite temperature: There, the order parameter of the transition is the spin, not the chirality, {\it i.e.\/}, no occurrence of the spin-chirality decoupling. In $d=1$ dimension, on the other hand, it has been known that the short-range Heisenberg SG exhibits only a $T=0$ transition both in the spin and in the chirality. Thus, the chiral-glass phase, if any, arises in intermediate dimensions around $d=3$. Indeed, numerical simulations for the Heisenberg SG in generalized dimensions, including $d=2$ \cite{KawaYone} and $d=4,5$ and $\infty$ dimensions \cite{ImaKawa}, have been done recently. Though useful information was obtained from these analyses,  intrinsic limitations also exist: {\it e.g.\/},  (i) the controlling parameter $d$ cannot be changed continuously so that a fine-tuning of the phenomena was impossible, and (ii) in higher dimensions, thermalization of larger systems became increasingly difficult due to the rapid increase of the total number of spins $N=L^d$, $L$ being the linear size of the lattice. 

In order to shed further light on the issue from somewhat different perspective, we consider here a different type of Heisenberg SG model, {\it i.e.\/}, 
the 1D Heisenberg SG model interacting via the long-range interaction which decays with distance as a power-law with an exponent $-\sigma$. For sufficiently small $\sigma$, the model is expected to reduce to an infinite-range MF model corresponding to $d=\infty$, while, for sufficiently large $\sigma$, the model is expected to reduce to the $d=1$ model with the short-range interaction. Hence, the variation of $\sigma$ in the 1D power-law SG model might mimic that of the dimensionality $d$ of the short-range SG model. Indeed, a recent numerical study on the corresponding 1D Ising SG model by Katzgraber and Young supported such correspondence \cite{Katz}.

Of particular interest here is whether the 1D Heisenberg SG with the long-range power-law interaction exhibits the spin-chirality decoupling for appropriate values of $\sigma$. 
In the present paper, we study by extensive Monte Carlo simulations the nature of both the spin and the chirality orderings of this model. 
%Such a calculation possesses several advantages: First, 1D models are easier to simulate than higher-dimensional counterparts which enables us to investigate larger lattice sizes. Second, the controlling parameter $\sigma$ can be varied continuously, making a fine-tuning of the phenomena possible in contrast to the case of the dimensionality. Third, certain analytical results are available in 1D, which could give us a useful guide in interpreting the numerical data. 

\section{Model}

The model we consider is the 1D classical Heisenberg model interacting via the random long-range interaction $J_{ij}$. The Hamiltonian is given by
   \begin{equation}
     H = - \sum_{<i,j>} J_{ij} \vec{S_{i}}\cdot \vec{S_{i}},
   \end{equation}
where $\vec S_i$ is the three-component classical Heisenberg spin variable, $\vec S_i=(S_i^x,S_i^y,S_i^z)$ with $|\vec S_i|=1$.
The interaction $J_{ij}$ is assumed to obey the Gaussian distribution, decaying  with distance $r_{ij}$ as a power-law,
   \begin{equation}
     J_{ij} = C \frac{\epsilon_{ij}}{r_{ij}^{\sigma}}, \ \ \ \ 
     C = \sqrt{\frac{N}{\sum_{i,j} r_{ij}^{-2 \sigma}}}.
   \end{equation}
where the $\epsilon _{ij}$ is chosen according to the Gaussian distribution with zero mean and the standard deviation unity:
   \begin{equation}
     {\cal P}(\epsilon_{ij}) = \frac{1}{\sqrt{2\pi}} \exp (-\epsilon_{ij}^2/2),
   \end{equation}
%
% while the normalization constant $C$ is given by
%
%   \begin{equation}
%     C = \sqrt{\frac{N}{\sum_{i,j} r_{ij}^{-2 \sigma}}}.
%   \end{equation}
%
In order to make the total energy extensive, the exponent $\sigma$ should be greater than $1/2$.

We impose periodic boundary conditions by placing the spins on a ring. Then, the distance between the $i$-th and the $j$-th spins $r_{ij}$ is given by \cite{Katz}
   \begin{equation}
     r_{ij} = \frac{L}{\pi} \sin \left( \frac{\pi |i-j|}{L} \right),
   \end{equation}
where $L$ is the total number of spins.

As mentioned, for sufficiently small and large $\sigma$, the model reduces to an infinite-range  $d=\infty$ model  and a short-range $d=1$  model, respectively. In the limit of $\sigma \rightarrow \infty$, in particular, the model should reduce to the nearest-neighbor model where frustration is totally irrelevant. Hence, one expects that the spin-chirality decoupling would arise neither in the small-$\sigma$ nor large-$\sigma$ limit: It could possibly arise only for intermediate values of $\sigma$. If one notes that the dimension $d=3$ probably lies close to the lower critical dimension of the SG order in the short-range Heisenberg SG model, the spin-chirality decoupling of the present 1D long-range SG model would arise, if any, near the borderline value of $\sigma$ separating the regions of a finite-temperature SG transition and a zero-temperature SG transition.

Thanks to its one-dimensionality, some analytical results are available for the 1D long-range SG model \cite{BMY,Kotliar,Chang}. On increasing $\sigma$ from $\sigma=0$, a finite-temperature SG transition changes its character from the MF one to the non-MF one beyond the borderline value of $\sigma$. This borderline value of $\sigma$, ``lower critical $\sigma$'',  is known to be $\sigma =\frac{2}{3}$. In the range of $\frac{2}{3} < \sigma < 1$, the SG transition still takes place at a finite temperature, but is governed by the non-MF long-range fixed point, characterized by an ``exact'' SG critical-point decay exponent $\eta_{{\rm SG}}=2-\sigma$. For $\sigma$ greater than  the ``upper critical $\sigma$'', $\sigma=1$, the SG transition occurs only at zero temperature with an exponent $\eta_{{\rm SG}}=1$, which is generically expected for any zero-temperature transition with the non-degenerate ground state. 

Previous analytic work did not consider the possibility of the spin-chirality decoupling \cite{BMY}. If one recalls the abovementioned $\sigma$-$d$ analogy, the spin-chirality decoupling might be expected for the range of $\sigma$ around the upper critical $\sigma$, $\sigma=1$. 

% in search for the signature of the spin-chirality decoupling. One should also note that, even in the case of a zero-temperature transition, the possibility of the spin-chirality decoupling still exists if the correlation-length exponents of the spin and the chirality associated with a $T=0$ transition differ with each other: The difference of the correlation-lengths exponents means an unusual situation that there are two diverging length scales, one associated with the spin and the other associated with the chirality, in this zero-temperature transition.  

\section{Monte Carlo simulations}

We perform an equilibrium MC simulation of the model. The power-law exponent $\sigma$ is set to $\sigma=1.1$, which lies in the region where we would expect the spin-chirality decoupling, if any. In our simulation, we make use of the temperature exchange MC method combined with the standard heat-bath updating. The lattice sizes studied are $L=64$, $128$, $256$, $512$ and $1024$, where the sample
average is taken over 128-512 independent bond realizations.  

The local chirality $\chi_{i \mu}$ at the $i$th site in the $\mu$
direction is defined by 
\begin{equation}
\chi_{i \mu} = \vec{S}_{i+\hat{e}_\mu} \cdot (\vec{S}_i \times \vec{S}_{i-\hat{e
}_\mu}), 
\label{scalar-chirality}
\end{equation}
$\hat{e}_\mu (\mu =x,y,z)$ being a unit lattice vector along the
$\mu$ axis. 

We probe the ordering of both the chirality and the spin by looking at the associated Binder ratios, {\it i.e.\/}, the spin Binder ratio $g_{{\rm SG}}$ and the chirality Binder ratio $g_{{\rm CG}}$, as  well as the associated finite-size correlation lengths, {\it i.e.\/}, the spin correlation length $\xi_{{\rm SG}}(L)$ and the chirality correlation length $\xi_{{\rm CG}}(L)$. Detailed definitions of these quantities have been given in Ref.\cite{HukuKawa2}. Both the Binder ratio $g$ and the dimensionless finite-size correlation length $\xi(L)/L$ have widely been used in numerical simulations in identifying the transition point. Since both quantities are dimensionless, the data for different size $L$ are expected to exhibit a crossing or a merging at a transition point.

In Fig.1, we show the size and temperature 
dependence of the Binder ratio $g_{{\rm SG}}$ (left) and of 
the dimensionless  correlation length $\xi_{{\rm SG}}(L)/L$ (right)
for the spin. As can be seen from the figure,
the spin Binder ratio decreases with increasing
$L$,
indicating the absence of a finite-temperature SG order. This is consistent with the result of analytical calculations showing $T_{{\rm SG}}=0$ for $\sigma>1$ \cite{BMY,Kotliar,Chang}. 

\begin{figure}[h]
\begin{center}
\includegraphics[scale=0.72]{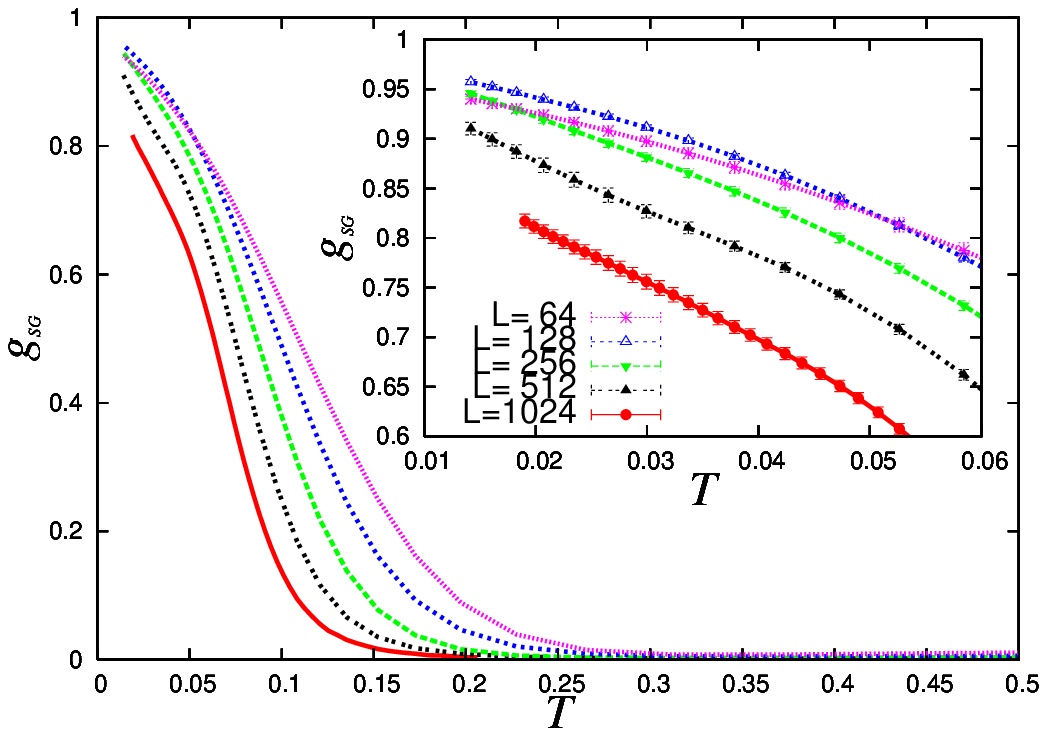}
\includegraphics[scale=0.72]{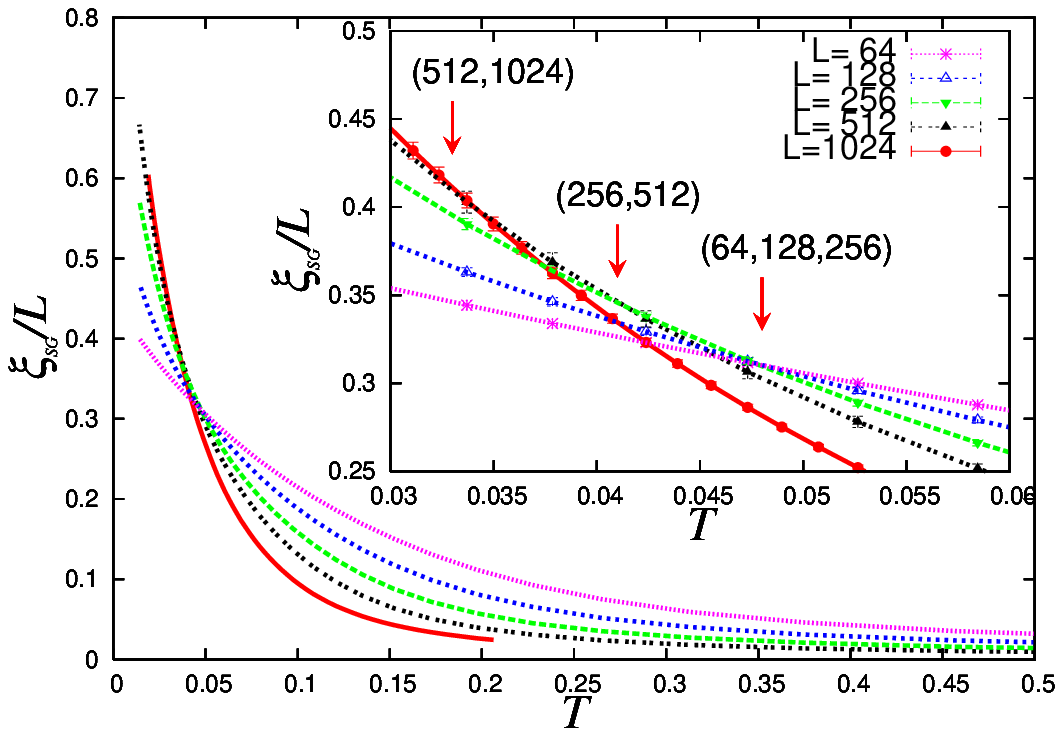}
\end{center}
\caption{
The temperature and size dependence of the spin Binder ratio (left) and of the dimensionless spin correlation length (right) for $\sigma=1.1$.
}
\end{figure}
%
%\begin{figure}[h]
%\begin{center}
%\includegraphics[scale=0.32,angle=-90]{1DHSG-Scorre.eps}
%\end{center}
%\caption{
%The temperature and size dependence of the dimensionless spin correlation lengt%h.
%}
%\end{figure}
%

By contrast, the behavior of the dimensionless correlation length points to the opposite at first glance, {\it i.e.\/}, the data for different $L$ appear to cross at a finite temperature  $T\simeq 0.05$ suggesting the occurrence of a finite-temperature SG transition: See the main panel. A closer inspection of the data,  however, has revealed that, although such a crossing indeed occurs at an almost size-independent temperature for smaller sizes $L\leq 256$, the crossing temperature rapidly shifts toward lower temperatures for larger sizes $L\geq 512$: See the inset.
Such a behavior is fully consistent with the occurrence of the spin-chirality decoupling occurring beyond the crossover length scale $L^*\simeq 500$.
Hence, the asymptotic behavior of the spin correlation length is eventually consistent with that of 
the Binder ratio and with the known analytical result.

%In order to see this asymptotic behavior, however, one needs to investigate fairly large lattices, say, $L\gsim 500$, particularly in the case of the correlation length. Otherwise, the data for the spin Binder ratio $g_{{\rm SG}}$ and of the spin correlation length $\xi_{{\rm SG}}/L$ give mutually conflicting answers: The former the absence of a SG transition, while the latter the existence of a finite-temperature SG transition at $T\simeq 0.05$ ! In the present case, the Binder ratio should be better trusted than the correlation length, since both the analytic result and the numerical result for larger sizes unambiguously indicate that there is no finite-temperature SG transition at this value of $\sigma$.

%
%
\begin{figure}[h]
\begin{center}
\includegraphics[scale=0.72]{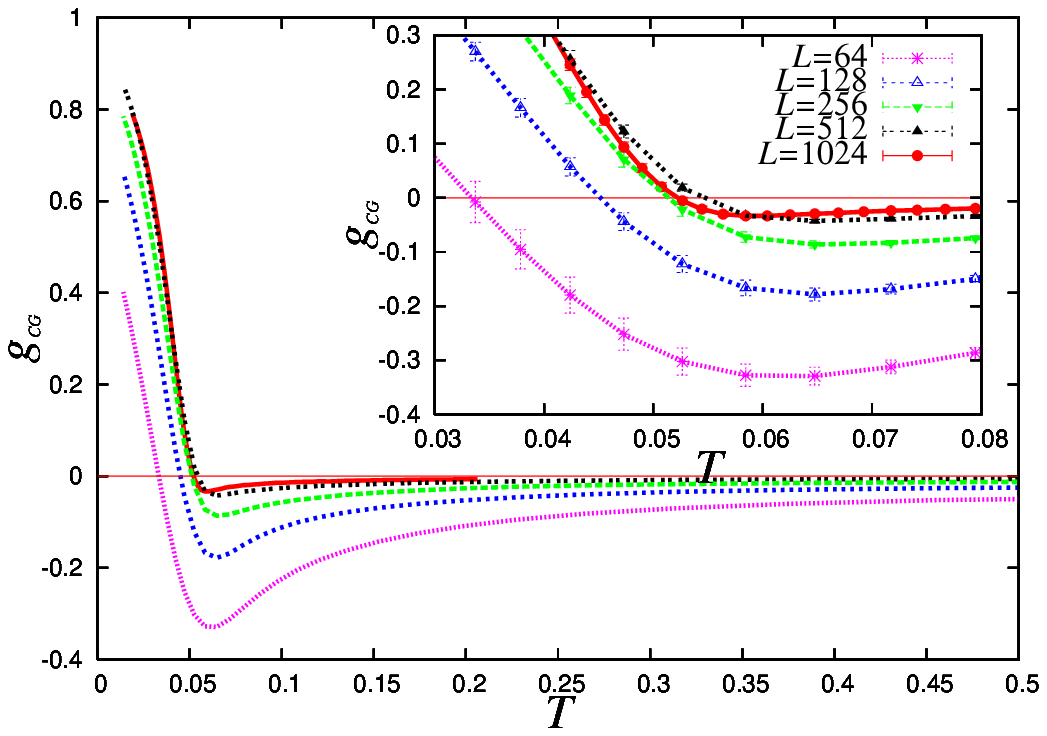}
\includegraphics[scale=0.72]{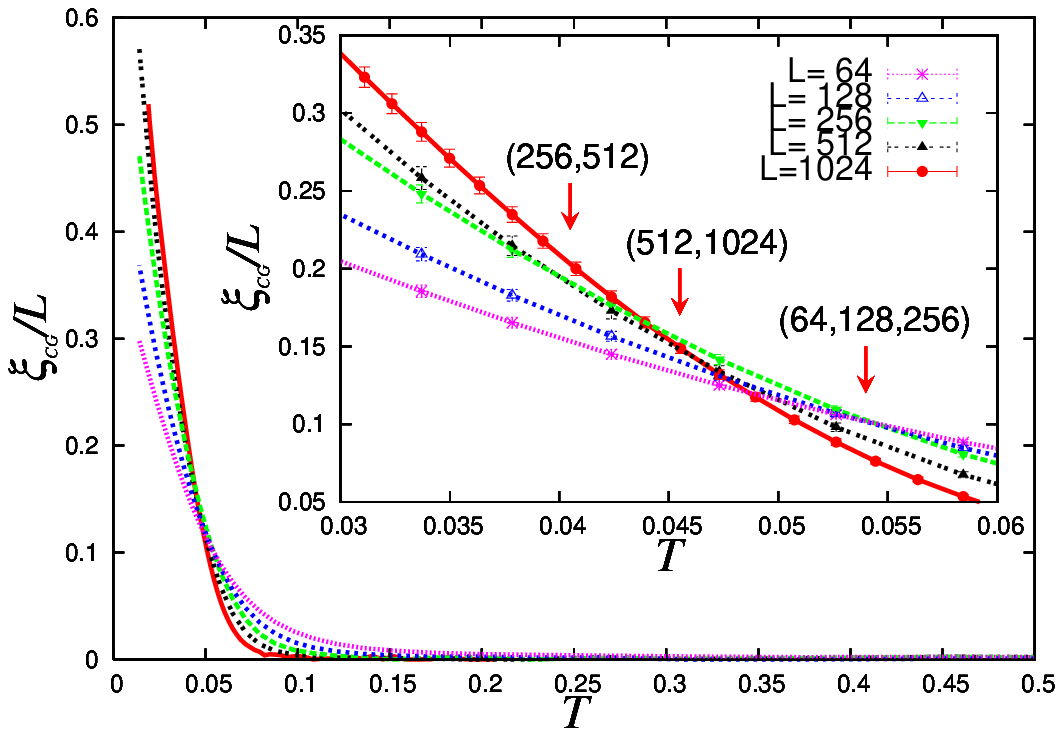}
\end{center}
\caption{
The temperature and size dependence of the chirality Binder ratio (left) and of the dimensionless chirality correlation length (right) for $\sigma=1.1$.
}
\end{figure}
%
%\begin{figure}[h]
%\begin{center}
%\includegraphics[scale=0.31,angle=-90]{1DHSG-Ccorre.eps}
%\end{center}
%\caption{
%The temperature and size dependence of the dimensionless chirality correlation %length.
%}
%\end{figure}
%
Next, we turn to the chirality ordering. In Fig.2,
we show the size and temperature 
dependence of the Binder ratio $g_{{\rm CG}}$ (left) and of 
the dimensionless correlation length $\xi_{{\rm CG}}(L)/L$ (right) of the chirality for the case of $\sigma=1.1$. As can be seen from the figure,
the Binder ratio exhibits a negative dip at a finite temperature $T=T_{dip}(L)$.
The temperature $T_{dip}(\infty)$ is expected to give a chiral-glass
transition temperature, which is estimated to be $T_{{\rm CG}}\simeq 
0.05$. This suggests the occurrence of 
a finite-temperature chiral-glass (CG) transition.
The dimensionless chirality correlation
length exhibits a behavior similar to that of the spin correlation length,
at least for the sizes $L\leq 512$. In sharp contrast to the
spin correlation length, however, the crossing temperature between our two largest
sizes $L=512$ and $L=1024$ now {\it shifts toward a higher temperature\/} 
as compared with the one between
$L=256$ and $L=512$, suggesting the occurrence of a finite-temperature
CG transition at $T_{{\rm CG}}\simeq 0.05$, consistently with the
estimate based on the chirality Binder ratio. Note that the occurrence of a finite-temperature {\it chiral\/}-glass transition at $\sigma=1.1$ does not contradict any known analytical result on this model.

The combined data for the spin and for the chirality give a fairly strong numerical support for the occurrence of the spin-chirality decoupling in the present model, {\it i.e.\/}, $T_{{\rm CG}}\simeq 0.05$ and $T_{{\rm SG}}=0$. This decoupling becomes clear only after studying larger lattices $L\gsim 500$, suggesting that the crossover length scale in this model might be rather large, $L^*\simeq 500$, presumably reflecting the long-range nature of the interaction.

\section{Summary and discussion}

By numerically investigating the spin and the chirality orderings of the one-dimensional Heisenberg SG with the long-range power-law interaction for the case of $\sigma =1.1$, we observed a strong numerical evidence that the CG transition occurs at a finite temperature while the standard SG transition occurs only at zero temperature, {\it i.e.\/}, the occurrence of the spin-chirality decoupling. If one believes the $\sigma - d$ analogy, our present observation may  give some  support to the occurrence of the spin-chirality decoupling in the original $d=3$-dimensional Heisenberg SG.

We wish to emphasize again that the spin-chirality decoupling has become eminent when one studies larger systems. This is particularly so when one looks at the correlation lengths: Remember that the data of correlation lengths for smaller lattices $L\lsim 500$ spuriously suggested the occurrence of a simultaneous spin and chiral transition at a finite temperature, {\it i.e.\/}, the absence of the spin-chirality decoupling, which, however, contradicted the analytical result on this model. Namely, if at $\sigma=1.1$ there were a simultaneous spin and chiral transition at a finite-temperature without the spin-chirality decoupling, the standard RG analysis should apply, inevitably yielding the exponent relation $\eta_{{\rm SG}}=2-\sigma =0.9$. Since the SG correlation function at finite $T_{{\rm SG}}$ decays with distance $r$ as $r^{-(\eta-1)}$, however, this leads to an immediate contradiction. Therefore, a simultaneous spin and chiral transition without the spin-chirality decoupling as apparently suggested from the correlation-lengths data for smaller sizes $L\lsim 500$ is not allowed at $\sigma=1.1$.

It is also a bit surprising that the spin Binder ratio and the normalized spin correlation length exhibit quite different behaviors for moderate lattice sizes $L\lsim 500$. While the absence of the standard SG order is evident in the spin Binder ratio already from rather small lattices, it becomes appreciable in the corresponding spin correlation length only for larger lattices, say, $L\gsim 500$. In this connection, it is sometimes mentioned in the literature that the correlation length might be the best quantity to look at in the study of phase transition, being superior to, {\it e.g.\/}, the Binder ratio \cite{LY}. However, our result indicates that this is not always the case: In the present occasion, on the contrary, the Binder ratio is a better quantity than the correlation length, at least for moderate lattice sizes. Of course, both quantities have given the same conclusion for large enough lattices, as it should be. 
%Anyway, our result gives us a warning that one should not overtrust the behavior of a single quantity, but rather, investigate the behaviors of {\it several independent quantities\/}, and carefully examine their consistency.

\end{document}